\documentclass[usenatbib]{mn2e}
\usepackage{times}
\usepackage{epsfig}
\usepackage{amsmath}
\title[The stellar mass content of distant galaxy groups]{The stellar mass content of distant galaxy groups}
\author[Balogh et al.]{Michael L. Balogh$^{1}$, Dave Wilman$^{2}$,
  Robert D. E. Henderson$^{1}$, Richard
  G. Bower$^{3}$, \newauthor
David Gilbank$^{4}$, Richard Whitaker$^{3}$, Simon L. Morris$^{3}$, George Hau$^{3}$, 
J.~S. Mulchaey$^{5}$, 
\newauthor A. Oemler Jr.$^{5}$ and R. G. Carlberg$^{4}$
\\
$^{1}$Department of Physics and Astronomy, University of Waterloo, Waterloo, Ontario, N2L 3G1, Canada\\
$^{2}$Max--Planck--Institut f{\" u}r extraterrestrische Physik, Giessenbachstrasse 85748 Garching Germany\\
$^{3}$Department of Physics, University of Durham, Durham, UK, DH1 3LE\\
$^{4}$Department of Astronomy and Astrophysics, University of Toronto,
Toronto, Ontario, M5S 3H8, Canada\\
$^{5}$Observatories of the Carnegie Institution of Washington, 813 Santa Barbara Street, Pasadena, California, USA\\
}
\date{\today}

\def\gtrsim{\mathrel{\raise0.35ex\hbox{$\scriptstyle >$}\kern-0.6em
\lower0.40ex\hbox{{$\scriptstyle \sim$}}}}
\def\lesssim{\mathrel{\raise0.35ex\hbox{$\scriptstyle <$}\kern-0.6em
\lower0.40ex\hbox{{$\scriptstyle \sim$}}}}

\def\kms{km~s$^{-1}$}
\def\keq{K^{0.4}_{\rm eq}}
\def\k04{K^{0.4}}
\def\s04{S1^{0.4}}
\def\mlk{$M_{200}/L_K$}
\def\mlopt{$M_{200}/L_{\rm opt}$}
\def\mls{$M_{\rm stellar}/L_K$}
\def\oii{[O{\sc ii}]}
\def\ewoii{$W_{\rm \circ}$(\oii)}
\begin{document} 
\maketitle
\begin{abstract}
We have obtained near-infrared imaging of 58 galaxy groups, in the
redshift range $0.1<z<0.6$, from the William Herschel Telescope and
from the {\it Spitzer} IRAC data archive.  The groups are selected from
the CNOC2 redshift survey, with additional spectroscopy from the Baade
telescope (Magellan).  Our group samples are statistically complete to
$K_{\rm Vega}=17.7$ (INGRID) and [3.6$\mu m$]$_{\rm AB}=19.9$ (IRAC).  
From these data we
construct near-infrared luminosity functions, for groups in bins of
velocity dispersion, up to 800 \kms, and redshift.
The total amount of near-infrared luminosity per group is compared with
the dynamical mass, estimated from the velocity dispersion, to compute
the mass-to-light ratio, \mlk.  We find that the \mlk values in these
groups are in good
agreement with those of their statistical descendants at $z=0$, with no evidence
for evolution beyond that expected for a passively evolving population.  There is a
trend of \mlk\ with group mass, which increases from \mlk$\approx 10$ for
groups with $\sigma<250~\mbox{\kms}$ to \mlk$\approx 100$ for
$425~\mbox{\kms}<\sigma<800~\mbox{\kms}$.  This trend is weaker, but still present,
if we estimate the total mass from weak lensing measurements.  In terms of stellar
mass, stars make up $\gtrsim 2$ per cent of the mass in the smallest
groups, and $\lesssim 1$ per cent in the most massive groups.  We also
use the near-infrared data to consider the correlations between
stellar populations and stellar masses, for group and field galaxies at
$0.1<z<0.6$.  We find that 
fewer group galaxies show strong \oii\ emission compared with
field galaxies of the same stellar mass and at the same redshift. We conclude that most of the
stellar mass in these groups was already in place by $z\sim 0.4$, with
little environment-driven evolution to the present day.
\end{abstract}
\section{Introduction}\label{sec-intro}
In hierarchical models of structure formation, small groups of galaxies
represent an important environment that connects active, star-forming field galaxies 
to quiescent cluster galaxies.  In this context they may be the
environment in which the star formation histories of galaxies are
dramatically altered.  Since a large fraction of galaxies in the
nearby Universe exist within groups
\citep[e.g.][]{HG82,Eke-groups}, such group-driven
transformations could have a dominant influence on galaxy evolution, at least since $z\sim 1$.
However, galaxy groups are not well understood relative
to field galaxies and larger clusters, especially beyond the local universe.  Their low-contrast relative to
the background means that deep, highly complete redshift surveys are
needed to compile large enough samples of member galaxies, and to
minimize the effects of selection biases.

Now that the global history of star formation over the past few billion
years is well established \citep[e.g.][]{Heavens,Hopkins04,Juneau+04}, the next important goal is to trace star
formation and efficiency in
different clustering environments \citep[e.g.][]{2dfsdss,Eke-groups2,DEEPII_envt}.
An important integral quantity that traces this evolution is the relative
amount of luminous matter in groups of different masses.  Studies at
optical wavelengths have shown that, locally at least, the ratio between
dynamical mass and optical light
(\mlopt) increases strongly with mass, up to systems with masses of a few
times $10^{13} M_\odot$, typical of poor clusters
\citep[e.g.][]{G+02,Eke-groups2}.  
Semi-analytic models of galaxy formation with reasonable feedback
schemes are in good agreement with this trend \citep{Eke-groups2}.
Recently, \cite{PHCH} have made measurements of 
\mlopt\ in galaxy groups selected from the second Canadian Network for 
Observational Cosmology (CNOC2) field galaxy redshift survey
\citep{CNOC_groups}, with masses determined by weak lensing 
\citep{Hoek01}.  They, too, find a trend of increasing \mlopt\ that is
consistent with that measured at lower redshift.

A difficulty in interpreting the above results is the
sensitivity of optical luminosity to recent star formation, which has a
strong environmental dependence \citep[e.g.][]{2dfsdss,Blanton04,BaldryV}.
Near-infrared (NIR) luminosities are superior in this respect, because they
trace stellar mass, independent of star formation history, to within a
factor of a few \citep[e.g.][]{BdJ_ML}.  A few near-infrared studies of
nearby groups have been undertaken \citep{LMS,RBGMR}, and these
observations generally find an increase in \mlk\ (the ratio between
dynamical-mass and near-infrared light) with mass, comparable to
that seen at optical wavelengths. However, until now such measurements have not
yet been reported for more distant groups.

It is the purpose of this paper to measure \mlk\ for
groups  at $0.1<z<0.6$ for the first time, using both
ground-based (K-band) and {\it Spitzer} (3.6$\mu m$) imaging of 58 galaxy groups from
the CNOC2 survey.
The paper is outlined as follows: In \S\ref{sec-data} below we review the CNOC2 survey, discuss 
our observations, data reduction techniques and methods of group member selection. We present 
the luminosity functions and \mlk\ as a function of group redshift and velocity
dispersion in \S\ref{sec-results}.  The implications of these
measurements, interpreted as stellar masses, are discussed in \S\ref{sec-discuss}, and our conclusions
are summarized in \S\ref{sec-conc}. Throughout this paper we assume a cosmology with matter 
density $\Omega_m=0.3$, energy density $\Omega_\Lambda=0.7$, and current Hubble constant 
$H_\circ=100h~\mbox{km~s}^{-1}~\mbox{Mpc}^{-1}$ with $h=0.75$.

\section{The Data}\label{sec-data}
\subsection{The redshift surveys }
Our galaxy group sample is derived from 
the CNOC2 redshift survey, which was conducted at the
Canada-France-Hawaii telescope using the multiobject spectrograph (MOS), with the primary goal of studying galaxy populations, 
clustering, and evolution \citep{CNOC2-I}. This spectroscopic and photometric survey 
consists of approximately 6000 galaxies with a measured redshift,
distributed in four widely separated patches covering a total of
$\sim1.5$ deg$^{2}$ of  
the sky.  Although the survey area is small relative to more recent
redshift surveys, the use of four noncontiguous patches ensures that
our group sample is not likely to be strongly biased due to cosmic
variance.  The 
survey includes \textit{U, B, V, $R_c$, I} photometry, and the initial spectroscopic sample was selected
from the $R_c$ (hereafter just $R$) images.  
The colours are k-corrected to the rest-frame using the 
template-fitting code of \citet{kcorrect}.  Each MOS pointing was
observed with two masks, to increase completeness in dense fields.
There are redshifts for about 50 per cent of galaxies to $R=21.5$; this
completeness is magnitude-dependent, as slits were primarily
allocated to brighter galaxies.

Recently we
re-observed $20$ fields in three of the four patches with the
LDSS2 spectrograph on Magellan \citep{CNOC2_groupsI}.  There were two main aims
of this additional spectroscopy.  The first was to cover each group
with sufficient masks (between one and three) to achieve near-100 per cent completeness at
bright magnitudes.  This is important because measurements of
quantities like total stellar mass will be strongly affected by the
brightest galaxies, and their small number per group will make
statistical corrections unreliable on an individual group basis.
Furthermore, the use of a larger telescope and better spectrograph
allows us to obtain redshifts for fainter galaxies, and we obtain a
good statistical sampling of galaxies as faint as $R=22$.  The main
advantage for our present purposes is that this increases the number of
members per group, allowing us to obtain better measurements of the
group geometry and velocity dispersion.  

From the CNOC2 and Magellan spectra we also measure the rest-frame
equivalent width of the \oii\ emission line, \ewoii, as an indicator of
ongoing star formation.  This line strength is measured following the
definition of \citet{B+97}.  In the
fields covered by our LDSS2 spectroscopy, the line measurements are the
same as those presented by \citet{CNOC2_groupsI}; for the rest of the CNOC2 sample,
the measurements will be presented in a future paper (Whitaker \&
Morris, in preparation).

\subsection{Galaxy Group Membership}\label{sec-ggm}
Virialized galaxy groups within the CNOC2 patches were initially identified in redshift space by 
\citet{CNOC_groups} using an iterative friends-of-friends algorithm on the 
catalogued galaxies. Over 200 groups were found, with an  
average of 3.8 confirmed members per group. The search by \citet{CNOC_groups} limited the sample's $k$-corrected 
and evolution-compensated absolute luminosity to brighter than
$M_R^{ke} = -18.5$ mag. 
To take advantage of the increased depth and completeness afforded us
by the Magellan data, in this paper we redefine the original group membership, in a manner
similar to that described in  
\citet{CNOC2_groupsI}.  To summarize briefly, an iterative procedure is
implemented to select galaxies with redshifts within twice the 
group velocity dispersion and with a transverse distance from the group
centre which is within $1/5$ of the line-of-sight distance.  In each
iteration, the velocity dispersion is recomputed using the Gapper
estimate \citep{Beers}, and the centre is recomputed as the
luminosity-weighted geometric centre of the group members.  This
procedure is applied to the whole group sample, including those that
were not 
re-observed with LDSS2, so that membership is consistently defined.  

To compute the virial  mass of all the galaxy groups we used a theoretical 
estimate based on the velocity dispersion, $\sigma$. 
Assuming that the groups are in 
dynamical equilibrium, the total mass contained within a volume of 
$\frac{4}{3} \pi R_{200}^{3}$ is given by
\begin{equation}\label{eqn-vm}
  \mathcal{M}_{200} = \frac{3}{G}R_{200}\sigma^{2},
\end{equation}
where the ``virial radius'' $R_{200}$ is 
\begin{equation}\label{eqn-r200}
R_{200}=\frac{\sqrt{3}\sigma}{10H_\circ\left(1+z\right)^{1.5}}.
\end{equation} 
Although we
are well aware that velocity dispersion is not a perfect indicator of
mass, especially for these poor systems with few members, we keep this
definition so that we can directly and easily compare our 
results with the literature \citep[e.g.][]{RBGMR}.    In
Section~\ref{sec-lens} we will consider how the use of available weak-lensing
masses affects our results.

As described in the following sections, we have obtained near-infrared
imaging for 58 groups, and these are the only groups we will consider in
the remainder of this paper.  The list of groups and their properties
are given in Table \ref{tab-groups}, for those groups with only data
from the original CNOC2 survey, and in Table~\ref{tab-groups2} for those
reobserved with Magellan.  Included in the tables are the
group coordinates, redshift, intrinsic velocity dispersion (i.e.,
accounting for velocity uncertainties), virial mass, and the number of group members 
we identify within $R_{200}$ above the $R-$ completeness limit (see
\S\ref{sec-zcomp}).   The velocity dispersions were computed
from all galaxies within $1/h$ Mpc of the group centre (unless there
were less than four members, in which case all galaxies were used); the
number of galaxies contributing to the measurement of $\sigma$ is given
in the table as $N_\sigma$, as is the the number of group members with either INGRID or
{\it Spitzer} observations, as described in Sections~\ref{sec-kband}
and~\ref{sec-spitzer}.
\begin{table*}
  \centering
    \caption{Properties of galaxy groups from the original \citet{CNOC_groups}
      sample (i.e. not reobserved at Magellan) with good
      near-infrared coverage.  Column 1 gives the group id from
      \citet{CNOC_groups}, modulus factors of 100 added to distinguish
      groups in different patches.  Columns 2-4 give the central position
      and redshift of the group.  Column 5 is the estimated intrinsic velocity
      dispersion, and column 6 is the number of galaxies used to obtain
      this number.  Columns 7 and 8 are the virial radius and mass
      calculated from the velocity dispersion.  Entries marked with an $^*$ are
      upper limits.  The numbers in columns (9), (10) and
      (11) are the number of galaxies with redshifts, within $R_{200}$
    in the $R-$band, {\it Spitzer} and INGRID catalogues,
    respectively.  }
    \label{tab-groups}
    \begin{tabular}{lcccccllccc} \hline \hline
(1) & (2) & (3) & (4) & (5) & (6) & (7) &(8) &(9)&(10)&(11)\\
  Group   & $\alpha$    & $\delta$   & $z$     & $\sigma$       & $N_\sigma$ &$R_{200}$  & $\mathcal{M}_{200}$      & $N_z$ & $N_z$    & $N_z$ \\
          & (J2000)     & (J2000)    &         & ($km~s^{-1}$)  & &
          $(kpc)$     & ($10^{12}\mathcal{M}_{\odot}$)  &  (R)     &
          ({\it Spitzer}) & (INGRID) \\\hline
1 & 222.4258728 & 9.050009727 & 0.1647 & $211\pm95$  & 8  &$388\pm174$ & $1.21\pm1.63$ & 5 & 5 & 4 \\ 
9 & 222.2241516 & 8.945289612 & 0.2616 & $194\pm61$  & 8  &$316\pm99$ & $0.83\pm0.78$ & 4 & 3 & 3 \\ 
11 & 222.2881775 & 8.830399513 & 0.2708 & $185\pm46$ & 12 &$298\pm75$ & $0.71\pm0.54$ & 4 & 4 & 0 \\ 
13 & 222.3945923 & 8.905659676 & 0.2712 & $397\pm68$ & 11 &$640\pm110$ & $7.05\pm3.62$ & 4 & 4 & 0 \\ 
15 & 222.1999817 & 8.963080406 & 0.3071 & $308\pm219$& 4  &$476\pm338$ & $3.14\pm6.71$ & 3 & 2 & 3 \\ 
16 & 222.5902405 & 9.103640556 & 0.3063 & $254\pm64$ & 5  &$392\pm100$ & $1.76\pm1.34$ & 4 & 4 & 0 \\ 
17 & 222.097641 & 8.784229279 & 0.3086 & $395\pm92$  & 8  &$610\pm141$ & $6.66\pm4.62$ & 3 & 0 & 3 \\ 
18 & 222.0555878 & 8.9292202 & 0.3234 & $261\pm264$  & 4  &$397\pm401$ & $1.89\pm5.74$ & 4 & 0 & 4 \\ 
19 & 222.5530853 & 8.960080147 & 0.3248 & $330\pm115$& 6  &$499\pm174$ & $3.79\pm3.96$ & 4 & 4 & 0 \\ 
21 & 222.4801636 & 9.646129608 & 0.3483 & $204\pm73$ & 7  &$301\pm107$ & $0.87\pm0.93$ & 6 & 0 & 5 \\ 
29 & 222.4464874 & 8.852370262 & 0.3736 & $309\pm158$& 7  &$443\pm227$ & $2.95\pm4.53$ & 4 & 4 & 0 \\ 
30 & 222.4998779 & 8.82020092 & 0.3938 & $335\pm87$ &  8  &$469\pm122$ & $3.67\pm2.85$ & 5 & 4 & 0 \\ 
36 & 222.3757324 & 9.153150558 & 0.4701 & $265\pm308$& 4  &$343\pm399$ & $1.68\pm5.85$ & 4 & 4 & 0 \\ 
202 & 36.58484268 & 0.4656900167 & 0.1883 & $341\pm288$& 5  &$607\pm513$ & $4.91\pm12.45$ & 4 & 4 & 0 \\ 
206 & 36.37388992 & 0.2616599798 & 0.2284 & $261\pm63$ & 11 &$442\pm106$ & $2.09\pm1.51$ & 6 & 4 & 0 \\ 
208 & 36.48381424 & 0.1982000023 & 0.2686 & $592\pm165$& 8  &$956\pm267$ & $23.38\pm19.54$ & 7 & 7 & 0 \\ 
217 & 36.44522095 & 0.3269200325 & 0.3082 & $696\pm276$ &5 &$1074\pm425$ & $36.31\pm43.12$ & 5 & 5 & 0 \\ 
221 & 36.52441406 & 0.07746999711 & 0.3578 & $73\pm95$ &6 &$107\pm139$ & $0.04\pm0.16$ & 1 & 1 & 0 \\ 
229 & 36.69626617 & 0.142960012 & 0.3833 & $540\pm139$ &13&$766\pm197$ & $15.56\pm12.02$ & 3 & 3 & 0 \\ 
234 & 36.5413208 & 0.4848300517 & 0.3974 & $284\pm111$ &7&$397\pm155$ & $2.24\pm2.62$ & 3 & 3 & 0 \\ 
239 & 36.36482239 & 0.2544400096 & 0.4083 & $608\pm96$ &6&$841\pm133$ & $21.72\pm10.29$ & 6 & 6 & 0 \\ 
301 & 141.086853 & 36.9005127 & 0.1072 & $66\pm86$ &6&$130\pm170$ & $0.04\pm0.15$ & 2 & 2 & 0 \\ 
302 & 140.2914429 & 36.71305847 & 0.112 & $123^*$ &3&$242^*$ & $0.26^*$ & 3 & 0 & 3 \\ 
303 & 141.0169373 & 37.0852623 & 0.1913 & $331\pm98$&7 &$587\pm175$ & $4.48\pm4$ & 8 & 6 & 0 \\ 
304 & 140.8880463 & 37.38243866 & 0.1914 & $121\pm159$ &3&$214\pm282$ & $0.22\pm0.86$ & 3 & 0 & 2 \\ 
305 & 140.9043732 & 36.91950989 & 0.2025 & $243\pm104$ &5&$425\pm182$ & $1.75\pm2.24$ & 3 & 3 & 0 \\ 
313 & 140.9319 & 37.01506042 & 0.2335 & $232\pm97$ &7&$391\pm164$ & $1.47\pm1.85$ & 4 & 4 & 0 \\ 
315 & 141.0404053 & 36.99619293 & 0.2435 & $243\pm45$ &13&$404\pm74$ & $1.66\pm0.92$ & 5 & 5 & 0 \\ 
317 & 140.9354401 & 37.34311676 & 0.2449 & $280\pm66$ &12&$465\pm109$ & $2.54\pm1.79$ & 9 & 0 & 6 \\ 
329 & 140.9833832 & 37.25820923 & 0.3223 & $487\pm79$ &4&$739\pm120$ & $12.22\pm5.94$ & 1 & 0 & 1 \\ 
333 & 140.8763885 & 37.14134979 & 0.3218 & $455\pm143$ &9&$692\pm217$ & $10.02\pm9.42$ & 6 & 5 & 4 \\ 
336 & 140.8822021 & 36.8615303 & 0.3629 & $463\pm531$ &4&$672\pm771$ & $10.05\pm34.58$ & 4 & 4 & 4 \\ 
337 & 140.9049683 & 37.7922287 & 0.3729 & $624\pm184$ &7&$895\pm264$ & $24.31\pm21.54$ & 6 & 0 & 4 \\ 
344 & 141.0652008 & 36.88687134 & 0.3735 & $277\pm77$ &12&$397\pm110$ & $2.12\pm1.77$ & 3 & 3 & 0 \\ 
346 & 141.004715 & 36.78276443 & 0.3733 & $393\pm30$ &26&$563\pm44$ & $6.06\pm1.41$ & 12 & 12 & 0 \\ 
348 & 140.3150482 & 36.71141052 & 0.3789 & $82\pm106$ &4&$117\pm151$ & $0.05\pm0.21$ & 1 & 0 & 1 \\ 
349 & 140.6838989 & 36.95294952 & 0.38 & $498\pm108$ &8&$710\pm154$ & $12.3\pm8.02$ & 3 & 0 & 3 \\ 
350 & 140.5788269 & 36.81409073 & 0.3783 & $285\pm85$ &7&$407\pm122$ & $2.31\pm2.08$ & 3 & 0 & 3 \\ 
351 & 140.965744 & 37.26145172 & 0.3795 & $143\pm88$ &6&$203\pm126$ & $0.29\pm0.53$ & 2 & 0 & 2 \\ 
352 & 141.0719757 & 36.94703293 & 0.3791 & $386\pm166$ &4&$550\pm236$ & $5.72\pm7.37$ & 3 & 3 & 0 \\ 
355 & 140.6422424 & 36.72232056 & 0.3907 & $349\pm220$ &5&$491\pm310$ & $4.18\pm7.9$ & 3 & 0 & 2 \\ 
357 & 140.3446808 & 36.62179947 & 0.3909 & $133\pm107$ &7&$187\pm151$ & $0.23\pm0.56$ & 2 & 0 & 2 \\ 
358 & 140.7977142 & 37.12773132 & 0.3908 & $287\pm53$ &11&$404\pm75$ & $2.32\pm1.29$ & 6 & 0 & 6 \\ 
359 & 140.9085388 & 37.121418 & 0.3911 & $266\pm127$ &4&$375\pm179$ & $1.85\pm2.65$ & 3 & 3 & 3 \\ 
361 & 140.7460632 & 36.83348846 & 0.4275 & $118\pm156$ &3&$159\pm211$ & $0.15\pm0.61$ & 2 & 0 & 2 \\ 
365 & 141.1464691 & 37.03060913 & 0.473 & $122\pm163$ &3&$158\pm211$ & $0.16\pm0.66$ & 3 & 3 & 0 \\ 
366 & 140.6099243 & 36.71389771 & 0.4728 & $455\pm191$ &9&$588\pm246$ & $8.49\pm10.68$ & 4 & 0 & 4 \\ 
\end{tabular}
\end{table*}

\begin{table*}
  \centering
    \caption{As Table~\ref{tab-groups}, but for the groups reobserved
    at Magellan \citep{CNOC2_groupsI}, with good NIR data.}
    \label{tab-groups2}
    \begin{tabular}{lcccccllccc} \hline \hline
(1) & (2) & (3) & (4) & (5) & (6) & (7) &(8) &(9)&(10)&(11)\\
  Group   & $\alpha$    & $\delta$   & $z$     & $\sigma$       & $N_\sigma$ &$R_{200}$  & $\mathcal{M}_{200}$      & $N_z$ & $N_z$    & $N_z$ \\
          & (J2000)     & (J2000)    &         & ($km~s^{-1}$)  & &
          $(kpc)$     & ($10^{12}\mathcal{M}_{\odot}$)  &  (R)     &
          ({\it Spitzer}) & (INGRID) \\\hline
23 & 222.3715668 & 9.511750221 & 0.3515 & $510\pm113$& 8  &$749\pm167$ & $13.58\pm9.06$ & 6 & 0 & 6 \\ 
24 & 222.2642517 & 9.116889954 & 0.3592 & $121^*$    & 10 &$177^*$ & $0.18^*$ & 6 & 5 & 0 \\ 
27 & 222.4210205 & 9.037179947 & 0.3725 & $181\pm148$& 3  &$260\pm212$ & $0.59\pm1.45$ & 3 & 2 & 3 \\ 
28 & 222.5953522 & 9.018850327 & 0.3728 & $161\pm79$ & 6  &$231\pm113$ & $0.42\pm0.61$ & 4 & 4 & 4 \\ 
31 & 222.3106995 & 9.188610077 & 0.3929 & $450\pm455$& 4  &$632\pm640$ & $8.94\pm27.14$ & 3 & 2 & 0 \\ 
32 & 222.4871063 & 8.929120064 & 0.3939 & $591\pm120$& 6  &$829\pm169$ & $20.22\pm12.32$ & 5 & 5 & 0 \\ 
33 & 222.3906708 & 9.480049133 & 0.4066 & $125\pm81$ & 6  &$173\pm113$ & $0.19\pm0.37$ & 2 & 0 & 2 \\ 
37 & 222.3853607 & 9.073459625 & 0.4713 & $236\pm79$ & 11 &$305\pm102$ & $1.18\pm1.19$ & 4 & 4 & 4 \\ 
38 & 222.3486328 & 8.980949402 & 0.5106 & $773\pm77$ & 14 &$962\pm95$ & $40.17\pm11.93$ & 10 & 8 & 0 \\ 
39 & 222.3684692 & 9.496768951 & 0.5366 & $461\pm104$& 13 &$559\pm126$ & $8.29\pm5.59$ & 8 & 0 & 8 \\ 
227 & 36.62685013 & 0.2123200148 & 0.3635 & $341\pm144$ &7&$495\pm209$ & $4.02\pm5.08$ & 4 & 4 & 0 \\ 
\end{tabular}
\end{table*}

\subsection{Redshift completeness}\label{sec-zcomp}
The CNOC2 redshift survey is $\sim$ 45 per cent complete on average,
with a dependence on $R$ magnitude.
For the original survey, \citet{CNOC2-I}
computed statistical weights that correct for this
incompleteness, and are valid to a limit of $R=21.5$.  The additional,
Magellan spectroscopy we obtained \citep{CNOC2_groupsI} improves the
depth around selected galaxy groups to $R=22$.  It is difficult to
combine the two samples in a statistically fair way; we thus
keep them separate, and will refer to them as the R21.5 and R22
samples, respectively.  These samples are exclusive: that is, galaxies
and groups in the R22 sample are removed from the R21.5 sample, even
though, of course, they formed part of the original CNOC2 survey.  The
statistical weights previously computed for both these 
surveys are magnitude- and position-dependent, but are computed from
ensemble averages that may not appropriately account for any
variation in completeness that depends on group richness, since groups
with more members may be more incompletely sampled, despite the
multiple-mask observing strategy.  We have
therefore recomputed the magnitude weights for group members as follows.
First, the group sample is divided by velocity dispersion into those
with $\sigma<200$ \kms, $200$\kms$<\sigma<500$\kms\ and $\sigma>500$ \kms.  For
each of these three subsamples, we compute the fraction of galaxies
within $R_{200}$ that have a redshift, as a function of $R$ magnitude
only, following the procedure described in \citet{CNOC2_groupsI}, but
omitting the radial weight.  For field galaxies, we
use a simple magnitude-dependent weight that is computed independently
for the R21.5 and R22 samples.  Despite all these efforts, our results
are only weakly dependent on the details of the weighting scheme.  In
particular, we find retrospectively that the incompleteness in the most
massive groups is at most $\sim 10$ per cent greater than the average.  Magnitude
weights are generally less than 1.5 for galaxies brighter than R=20,
and increase up to a maximum of 5 (i.e. 20 per cent completeness) at
R=22.  

\subsection{K-band Observations and Data Reduction}\label{sec-kband}
Over the course of three nights during March 9-13, 2001, 25 fields
centred on groups in the 9h and 14h CNOC2 patches were observed with the Isaac 
Newton Group Red Imaging Device (INGRID) on the $4.2 m$ William Herschel Telescope (WHT). 
Each field was observed using a nine-point dither
pattern with exposures of 10--15s in $K_s$.  The total
integration for each field varied, but was typically about 12 minutes.
The data were reduced using the {\sc ipipe} NIR reduction pipeline
\citep{Gilbank03}.  Briefly, this involved bad pixel-masking, flat fielding
and sky-subtraction using a local flat field, and offset-finding and
stacking of the sky-subtracted images to create a first-pass mosaic.
An object mask was generated using {\sc SExtractor}  \citep{sextractor}
to detect
objects in this mosaic, and the sky-subtraction process was repeated after  
applying this mask to the individual exposures.  This two step sky-subtraction
technique avoids over-subtraction of the sky, as faint objects
undetected in a single frame can otherwise lead to an overestimate of
the sky level.  The  second pass sky-subtracted frames were then
combined into a final mosaic using a 3-$\sigma$ clipped
mean.  

Our magnitude zeropoints (on the Vega system) and astrometric solution were calibrated using the
Two Micron All Sky Survey \citep[2MASS; see][]{2MASS}
point-source catalog $K_s$-band 4\arcsec\ standard aperture magnitudes. Although the extended source catalog 
would have been a preferable source of photometry, there were not enough matching objects 
in our fields to obtain a reliable calibration. We used
\textsc{SExtractor} v2.3.2
to obtain 4\arcsec\ aperture magnitudes from our observations, and
these were then compared to the matching 2MASS point-source catalog object magnitudes. 
For every field, the median of the differences between the \textsc{SExtractor} and 2MASS 
magnitudes was taken to be the zeropoint shift\footnote{In practice this was
done using the brightest sources, but fainter objects were included if
there were enough of them to overcome the larger uncertainties on their
individual magnitudes.}.

For our analysis  we adopt the \textsc{mag\_best} photometry 
output from \textsc{SExtractor}. This gives a \cite{K80} magnitude if there is no danger 
of a nearby object biasing the magnitude measurement by more than 10 percent. If such crowding is 
an issue, a corrected isophotal magnitude is used instead.
The perimeter of all images were trimmed such that most of the underexposed regions were not included 
in our analysis. 
We discarded all detected objects with \textsc{SExtractor} flags
greater than 8. Two of the 25 images obtained were rejected entirely; one due to poor image quality, 
and the other due to poor seeing ($\sim1.7\arcsec$). 
Objects were then matched with the CNOC2 photometric catalogues
\citep{CNOC2-I}.  Unfortunately, the global astrometry of the optical
catalogues is not better than $\sim 5$\arcsec.  This effect
was mitigated by adjusting the centroid of each field independently;
over the relatively small INGRID field-of-view ($4.1\arcmin$) the astrometry was good
enough to allow reliable matching.

\begin{figure}
\leavevmode \epsfysize=8cm \epsfbox{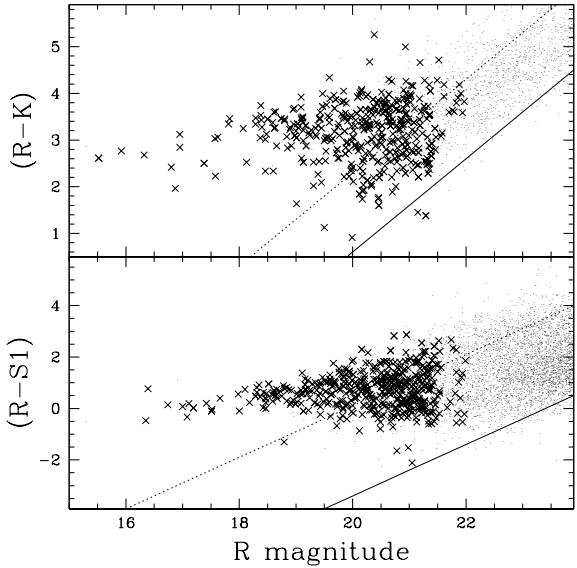} 
\caption{The optical-NIR colours of galaxies in the INGRID sample (top
  panel) and (for clarity) one-third of the {\it Spitzer} sample (bottom panel), as a function of $R$
  magnitude.  Crosses represent galaxies with spectroscopic
  redshifts. The solid lines represent the photometric completeness of 
  the IR data, which are $K=19.4$ and $S1=23.4$, respectively.  The
  dotted lines represent the limits of $K=17.7$ and $S1=19.9$ that
  correspond to the spectroscopic completeness limit for $z<0.6$ (see \S~\ref{sec-comp}).
\label{fig-checkIRcomp2}}
\end{figure}

In the top panel of Figure~\ref{fig-checkIRcomp2}, we show the K-R
colour as a function of R magnitude, for all objects classified as
galaxies.  From the absence of blue galaxies fainter than $R\sim 20$ we
determine that the  photometric magnitude limit is approximately
$K=19.4$.  We will show 
below (\S~\ref{sec-comp}) that this is much fainter
than the completeness of the spectroscopic sample (shown as the dotted
line), so all galaxies with redshifts are well-detected in the near-infrared.
 
\subsection{Spitzer observations}\label{sec-spitzer}
Data taken with the {\it Spitzer} telescope Infrared Array Camera
(IRAC) were obtained from the archive (GO program 64,
PI G. Fazio).  The
pipeline-reduced images were used, and objects were detected in IRAC
band 1 (AB magnitudes at 3.6$\mu$m, hereafter denoted S1) with {\sc SExtractor} v2.4.4.  The
detection criterion was a minimum of four pixels above a threshold of
1.5 times the global, rms background, with no filtering.  The
deblending algorithm used a minimum contrast of $0.00002$, with 64
sub-thresholds. These parameters were chosen to optimize the deblending
for matching with the optical catalogues.
The object fluxes were determined from the improved \textsc{mag\_auto}
magnitude, which is recommended for the most recent versions of {\sc SExtractor}.

Again, the relatively poor astrometry of the optical catalogues causes
some difficulty in cross-correlation.  For the 14h patch, we have
improved the astrometric solution based on several {\it Hubble Space Telescope}
pointings.  However, in the 2h and 9h fields, the astrometric
solution varies by $\sim 5$\arcsec\ across the overlapping IRAC
field-of-view.  Thus a larger matching radius was chosen ($4$\arcsec\
in the 2h field and $5$\arcsec\ in the 9h field), which causes an estimated $\sim 10$
per cent of sources to be incorrectly matched.
The photometric completeness is approximately S1=23.4, as shown in the bottom panel
of Figure~\ref{fig-checkIRcomp2}.  As is the case with the K-band data, this
is several magnitudes below our spectroscopic limit.

\subsection{Completeness}\label{sec-comp}
\begin{figure}
\leavevmode \epsfysize=8cm \epsfbox{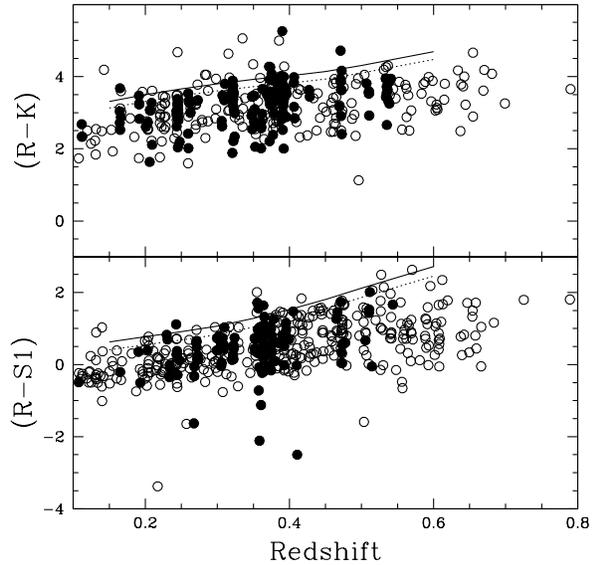} 
\caption{Observed optical-IR colours are shown as a function of
  redshift, for all galaxies with spectroscopic redshifts.  Filled
  symbols represent group members. The R band magnitudes come from the
  CNOC2 catalogue, while the S1 magnitudes (bottom panel) are from the
  {\it Spitzer} IRAC data, and the K magnitudes (top panel) are from INGRID.
  The lines are \citet{BC03} models for a solar metallicity, 13.7 Gyr
  model.  The dotted line is dust-free, while the solid line 
  includes $\tau_v=1$ magnitude of extinction, and thus provides a good
  estimate for the maximum possible colour for most normal galaxies.  
\label{fig-checkIRz}}
\end{figure}

We need to define the magnitude limits in both $K$ and $S1$ such that
the spectroscopic sample of galaxies above these limits is
statistically complete. This will be the case if $R<21.5$ (for the R21.5
sample) or $R<22$ (for the R22 sample) for all galaxies brighter than
these limits.  
Figure~\ref{fig-checkIRz} shows the optical-NIR colours as a
function of redshift, for the INGRID and {\it Spitzer} samples.  The solid
line shows a model of a solar metallicity, dust-reddened, 13.7 Gyr old stellar
population from the \citet{BC03} models.  Here and throughout the
paper, we use the two-component dust model of \citet{CF00}, assuming a
visible optical depth to the youngest stars (with age $<10^7$ yr) of 
$\tau_v=1$, and 30 per cent of this value toward older stars.  The
model provides a reasonable match to the red envelope 
of the data.  The data and model indicate that, at $z\leq0.6$, all
normal galaxies should have $(R-S1)\lesssim2.1$ and $(R-K)\lesssim4.3$.  Galaxies redder
than this (of which there are many in our photometric sample) are
almost certainly at higher redshift.  Given the R-completeness limits
of the R21.5 and R22 samples, we therefore have the following NIR
limits:
\begin{equation}
K_{\rm lim}=
\begin{cases}
17.2&R21.5\\
17.7&R22\\
\end{cases}
\end{equation}
\begin{equation}
S1_{\rm lim}=
\begin{cases}
19.4&R21.5\\
19.9&R22\\
\end{cases}
\end{equation}
In all cases, these magnitudes are well above the photometric
completeness limits.  Therefore, applying the weights appropriate to
the R21.5 or R22 sample will provide statistically complete samples to
these limits.

To compute the stellar luminosity of all group members, we need finally
to account for the fact that some groups are only partially covered by
{\it Spitzer} or INGRID imaging.  For {\it Spitzer}, this generally occurs for groups that
lie near the edge of the IRAC image.  The INGRID images were centred on
particular groups, but could still be incomplete if the virial
radius (Equation~\ref{eqn-r200}) is larger than the field of view, or if
there are background (or foreground) groups in the image that are not
as well centred.  For each group in the sample, therefore, we compute
how many of the spectroscopic members have NIR data from either set of
observations.  Groups are only included in the sample if the fraction of
members with NIR data is at least 2/3, and for each group the
members are weighted by the inverse of this fraction.  These groups are
listed in Table~\ref{tab-groups}, with the number of spectroscopically
confirmed group members with either INGRID or {\it Spitzer}
observations.  There are 29 (38)
groups with good INGRID (IRAC) coverage, and a total of 58 that
have coverage with at least one of the two instruments. When a galaxy
is observed with both instruments, we use the IRAC data since it has
the more stable zeropoint.

\subsection{Combined infrared sample}
\begin{figure}
\leavevmode \epsfysize=8cm \epsfbox{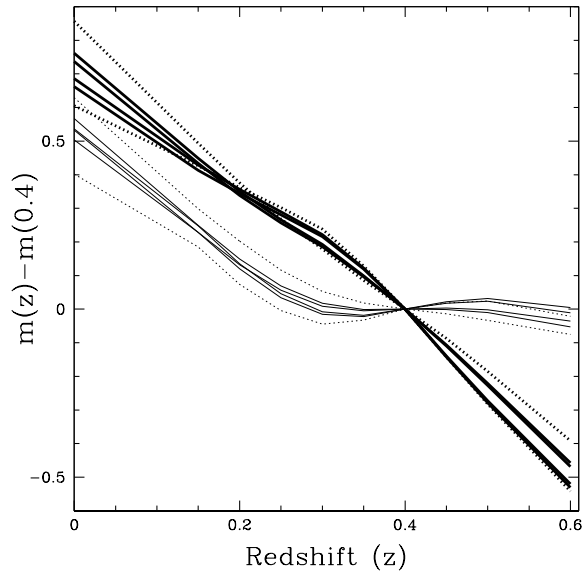} 
\caption{Model k-corrections, from a suite of \citet{BC03} models.  The
  thick lines correspond to IRAC S1 AB magnitudes, while the thinner
  lines represent K (Vega) magnitudes.  The solid lines, which are nearly
  indistinguishable from one another, represent solar metallicity
  models, with a Salpeter IMF. 
  The four lines are all 13 Gyr models, with a) single stellar
  population (ssp), no dust; b) ssp,
  $\tau_v=1$ extinction; c) constant star formation rate (SFR), no dust; and d) constant SFR,
  $\tau_v=1$ extinction.  The two models with dashed lines are constant
  SFR models, at supersolar and subsolar metallicity; these models
  provide the most different k-corrections.  There is almost no
  sensitivity to population age or initial mass function (not shown).
\label{fig-kcorrect}}
\end{figure}
To present luminosity functions it will be
useful to have all the infrared data (taken in two different filters
and over a range of redshifts) corrected to the same rest-frame
wavelength range.  We will take the approach, described in detail below, of
converting the {\it Spitzer} S1 AB magnitudes to equivalent $K-$band
(Vega) at $z=0.4$, for
ease of comparison to the literature.  This magnitude will be denoted
$\keq$ and is on the Vega system; note that $K_{AB}\approx K_{\rm Vega}+2.0$.

First, k-corrections are computed to convert the INGRID and IRAC
magnitudes separately to the corresponding observed wavelength range at
a fiducial redshift. To keep the correction as small as
possible, where convenient we will show the data k-corrected to
they typical group redshift of $z=0.4$, rather than $z=0$.  
To make these corrections we compute the difference in $K$ magnitude
relative to $z=0.4$ for a variety of 
\citet{BC03} galaxy models.  The results are shown in
Figure~\ref{fig-kcorrect}, and the model parameters are described in
the caption.  Since the
NIR fluxes are relatively insensitive to these parameters,
the choice is not critical, and to a good approximation
we find we can use 
\begin{equation}\label{eqn-kcorrk}
K(z)-\k04=
\begin{cases}
0& 0.3\leq z<0.6\\
-1.73z+0.52& z<0.3\\
\end{cases}
\end{equation}
\begin{equation}
S1(z)-\s04=
\begin{cases}
-2.50z+1.0& 0.4\leq z<0.6\\
-1.8z+0.72& z<0.4\\
\end{cases}
\end{equation}
where $\k04$ and $\s04$ refer to the observed band at $z=0.4$.
\begin{figure}
\leavevmode \epsfysize=8cm \epsfbox{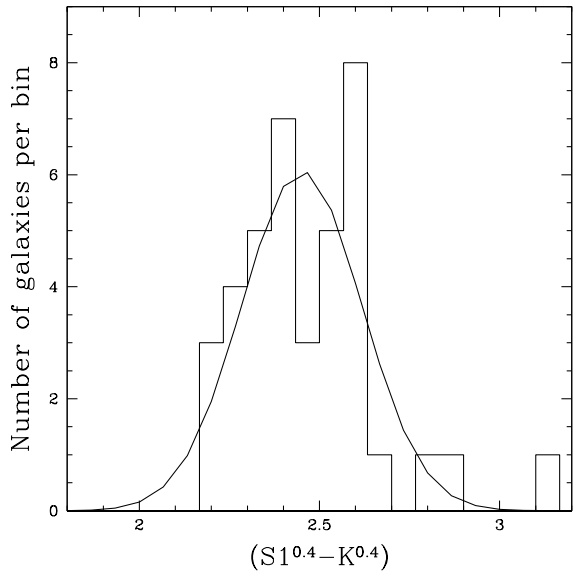} 
\caption{The distribution of $(\s04-\k04)$ colours for the 39 galaxies
  observed with both INGRID and IRAC, and with
  $0.3<z<0.6$, $K<17$ and $S1<20$. This colour is 
  insensitive to the nature of the stellar population, so we use the mean colour (2.45) as a
  conversion between the two magnitudes.  The solid line is a Gaussian
  with this mean, and standard deviation 0.17 mag, a best fit to the
  3-$\sigma$ clipped data.
\label{fig-cfSI3}}
\end{figure}
\begin{table*}
  \centering
    \caption{Average group properties, divided into redshift (column 1)
      and
      velocity dispersion ($\sigma$, column 2) bins.  The other columns
    are: (3) The number of groups contributing to each bin; (4) the
    total observed, statistically weighted, rest-frame K-band
    luminosity within $R_{200}$ and brighter than $M^\ast$; (5) The total rest-frame K-band
    luminosity including a correction for galaxies below $M^\ast$, assuming $\alpha=-0.8$; (6) the same as (5), but for
    $\alpha=-1.09$; (7) the average dynamical mass in this bin; (8), the
    mass-to-light ratio, defined as
    column (7) divided by column (5).}
    \label{tab-ltot}
    \begin{tabular}{cccccccc} \hline \hline
(1) & (2) & (3) & (4) & (5) & (6) & (7) & (8)\\ 
  Redshift&$\sigma$    & $N_{groups}$& $L_{K}$, $M_K<M_K^\ast$ &   $L_{K,tot}$  ($\alpha=-0.8$) &  $L_{K,tot}$ ($\alpha=-1.09$) &   $M_{200}$ & $M_{200}/L_{K,tot}$\\
   & (\kms)      &             &($10^{11} L_{K,\odot}$)           &             ($10^{11} L_{K,\odot}$)  &  ($10^{11} L_{K,\odot}$)   & ($10^{13} M_\odot)$& ($M_{\odot}/L_{K,\odot})$\\
\hline
0.1--0.25 & 0--250 & 7  & $ 2.03\pm0.4$ & $4.48\pm0.8$ & $6.17\pm1.1$ &   1.06 & $23.77\pm4.2$\\
          & 250--425 & 4  & $ 3.08\pm0.7$ & $6.8\pm1.6$ & $9.36\pm2.2$ &   3.25 & $47.79\pm11.4$\\
\hline
0.25--0.37 & 0--250 & 5  & $ 4.23\pm1.5$ & $9.33\pm3.4$ & $12.85\pm4.7$ &   0.63 & $6.79\pm2.5$\\
           & 250--425 & 7  & $ 4.96\pm1.5$ & $10.94\pm3.4$ & $15.07\pm4.6$ &   4.04 & $36.91\pm11.4$\\
           & 425--700 & 7  & $ 3.41\pm1.2$ & $7.52\pm2.6$ & $10.36\pm3.6$ &   18.04 & $239.88\pm84.4$\\
\hline
0.37--0.6 & 0--250 & 9  & $ 2.97\pm1$ & $6.55\pm2.3$ & $9.02\pm3.1$ &   0.39 & $5.99\pm2.1$\\
          & 250--425 & 11  & $ 6.87\pm0.8$ & $15.17\pm1.7$ & $20.9\pm2.3$ &   3.64 & $24.01\pm2.6$\\
          & 425--800 & 9  & $ 9.19\pm2.2$ & $20.28\pm4.9$ & $27.93\pm6.7$ &   21.69 & $106.95\pm25.8$\\
\hline\hline
0.1--0.6 & 0--250 & 21  & $ 2.95\pm0.6$ & $6.52\pm1.4$ & $8.98\pm1.9$ &   0.75 & $11.55\pm2.5$\\
         & 250--425 & 22  & $ 5.57\pm1$ & $12.3\pm2.2$ & $16.95\pm3.1$ &   3.67 & $29.81\pm5.4$\\
         & 425--800 & 15  & $ 7.1\pm1.4$ & $15.68\pm3.2$ & $21.59\pm4.4$ &   20.37 & $129.92\pm26.4$\\
\hline
    \end{tabular}
\end{table*}

After applying these corrections we can compare the $\k04$ and $\s04$
magnitudes for galaxies observed with both instruments.  In particular
there are 39 such galaxies, with $0.3<z<0.6$, $K<17$ and $S1<20$ (one
magnitude brighter than the  photometric completeness limit, to reduce
the impact of magnitude uncertainties).  The
distribution of ($\s04$-$\k04$) is shown for these galaxies in Figure~\ref{fig-cfSI3}.  After
clipping the only 3-$\sigma$ outlier, the mean and median are both 2.45,
and the standard deviation is 0.17 mag\footnote{
We note that this colour is also in excellent agreement with
predictions of the same models used to calculate the k-corrections: all
models considered predict $(\s04-\k04)=2.4\pm0.1$.}.  There is no significant trend
of this difference with rest-frame optical colour or redshift.  Thus we
take the dispersion to represent the uncertainty in photometric
calibration, including k-corrections and aperture effects.  We
convert our $\s04$ magnitudes to an equivalent $\k04$ by 
\begin{equation}
\keq=\s04-2.45.
\end{equation}
Finally, from Equation~\ref{eqn-kcorrk} we see that the rest-frame
K-band magnitudes are well approximated as 
\begin{align}
K_{\rm rest}&=\k04+0.52\\ \notag
            &=\keq+0.52.
\end{align} 
This k-correction between $z=0.4$ and $z=0$ is good to within about
0.05 mag, for all solar metallicity models considered.

\section{Results}\label{sec-results}
\begin{figure}
\leavevmode \epsfysize=8cm \epsfbox{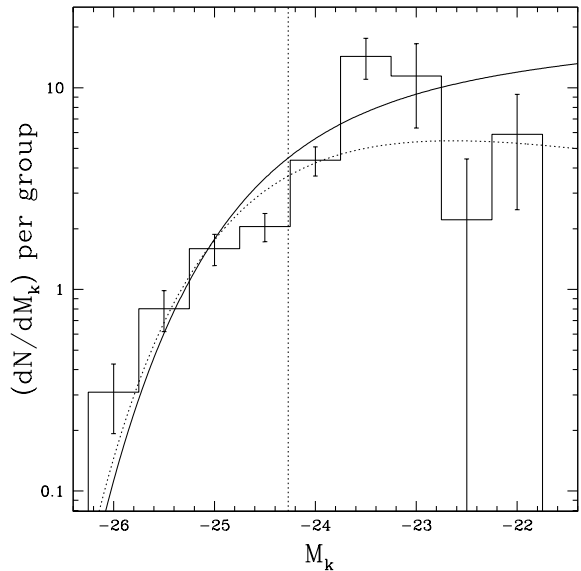} 
\caption{The luminosity function, per group, for all groups in the
  sample. The error
  bars are computed assuming Poisson statistics on the number of
  (unweighted) galaxies in each bin.  The vertical, dotted line
  represents the brightest magnitude limit of all the groups in
  the sample; to the right of that line the data are still statistically
  complete, but there are progressively fewer contributing groups.  The
  x-axis is the absolute rest-frame magnitude in the K filter.  The
  curved, solid (dashed) line is a Schechter function with $M_k^\ast=-24.38$ and
  $\alpha=-1.09 (-0.8)$.  
\label{fig-comblf}}
\end{figure}
\subsection{Luminosity Function}\label{sec-lfunc}

\begin{figure*}
\leavevmode \epsfysize=14cm \epsfbox{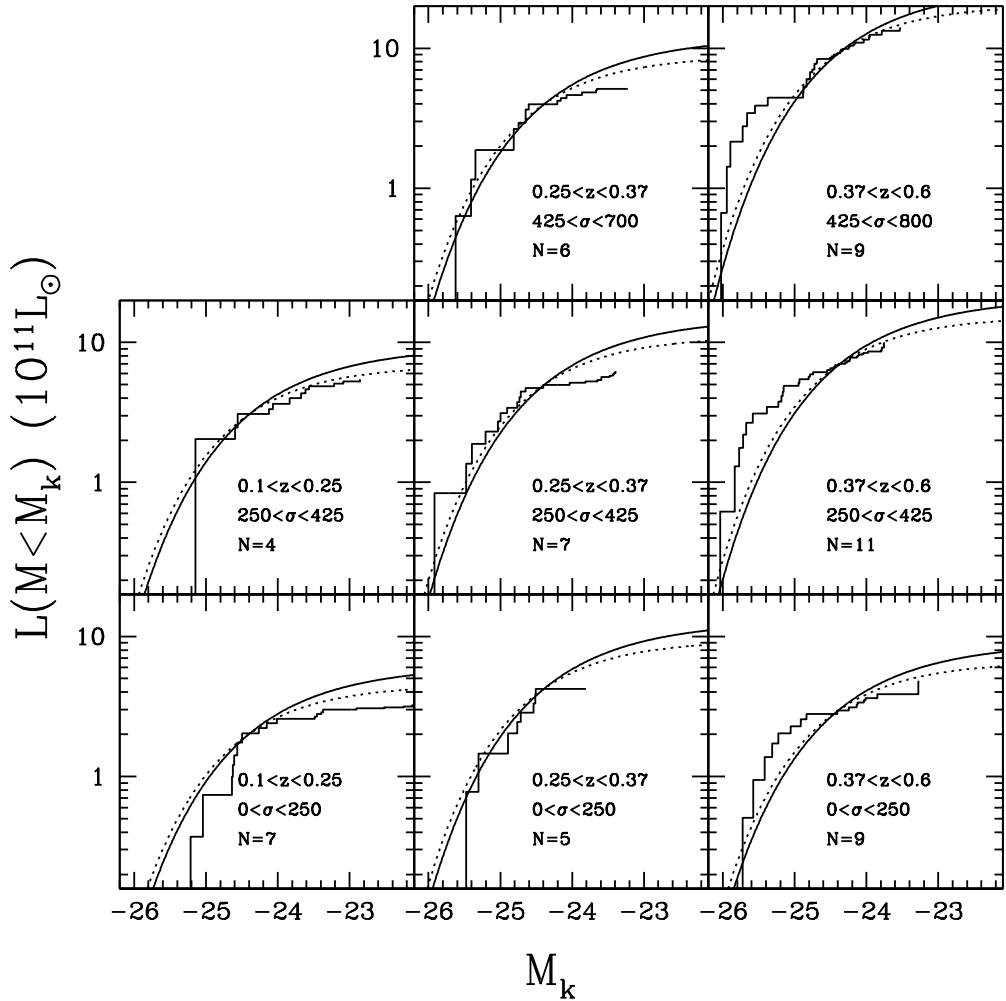} 
\caption{The cumulative luminosity per group, binned by group redshift
  and velocity dispersion as shown. Note that in this representation,
  the observational error bars (not shown) are not independent, and are
  largest at the brightest magnitudes.  The number of groups contributing
  to each redshift and velocity dispersion bin is also shown in each
  panel.  The
  x-axis is the rest-frame K-band absolute magnitude.  The
  curved, solid (dotted) line is a Schechter function with $M_k^\ast=-24.38$ and
  $\alpha=-1.09 (-0.8)$.  Both functions are normalized so that the total luminosity in galaxies
  brighter than $M_k^\ast$ matches the observations.  
\label{fig-comblf2}}
\end{figure*}
Most of our groups have less than about ten members with redshifts. To
construct a luminosity function it is therefore necessary to stack the
groups\footnote{As discussed in Section~\ref{sec-zcomp}, the spectroscopic
completeness is not a strong function of group size, which ensures that
this stacked group is not dominated by sparsely-populated groups with
large statistical weights; the weights are primarily a function only of
apparent magnitude.}.
When computing the luminosity functions we 
consider only galaxies within $R_{200}$, which itself is estimated from the
velocity dispersion of the group (Equation~\ref{eqn-r200}).  
Absolute magnitudes are computed from the $K_{\rm rest}$ magnitudes
described above, according to our adopted cosmological model.

Since our data combine two spectroscopic samples, with different
completeness limits, we first compute the luminosity function
separately for the R21.5 and R22 samples, and then combine them,
appropriately weighted according to the relative number of contributing
galaxies.  We have verified that the luminosity function for either
sample alone is statistically consistent with that of the combined
data.  Since the R22 sample is more complete, this suggests that any
systematic effects resulting from undersampling are masked by the
larger, statistical uncertainties. 

The luminosity
function for all groups in the survey
is shown in Figure~\ref{fig-comblf}.   The number of galaxies per magnitude bin are
weighted by the appropriate statistical weights, and divided 
by the number of groups contributing to that bin.  The dotted line
shows the luminosity corresponding to the magnitude limit of the
highest redshift group in the sample; at fainter luminosities, fewer
groups contribute to each bin.   

We will not attempt to derive Schechter function parameters from our data, which are
not much deeper than $M^\ast$ and include few spectroscopic group
members when the groups are binned by velocity dispersion (see
below).  However, it is necessary to assume some shape to correct the
total group luminosity for galaxies below the magnitude limit.  
\citet{Drory+03} have measured the field K-band luminosity functions out
to $z=1$, and found (for $h=0.7$) $M_k^\ast=-24.16-0.53z$;  at $z=0.4$,
therefore, $M_k^\ast=-24.38$.  We will adopt this value throughout the
paper, although we might expect $M^\ast$ to vary by $\sim 0.2$ mag over
the full redshift range of our sample. \citet{Drory+03}
cannot constrain the faint end slope, and so adopt $\alpha=-1.09$ measured from
2MASS \citep{Kochanek-KLF}.  However, the slope may be somewhat
shallower in dense environments, more like 
$\alpha=-0.8$ \citep[e.g.][]{IRLF,RBGMR}, so we will show both. 
The Schechter function using these parameters is shown in
Figure~\ref{fig-comblf}, 
normalized to match the total luminosity of galaxies brighter than $M_K^\ast$;
both provide a reasonable description of the data, within the uncertainties.

\subsection{Total group luminosities}\label{sec-lgroup}
In Figure~\ref{fig-comblf2} we show the {\it cumulative} luminosity
distribution in groups, divided into
redshift and group velocity dispersion bins.
This alternative
representation of the luminosity function shows the integrated
K-band luminosity per group, brighter than a given magnitude.  Models
are shown based on the Schechter functions presented in the previous section
($M_k^\ast=-24.38$ and $\alpha=-1.09$ or $\alpha=-0.8$), always renormalized to match the
total luminosity of all galaxies brighter than $M_k^\ast$.  There is no
dramatic change in the shape of the distribution with redshift or
velocity dispersion, although an evolution of $M_k^\ast$ of
$\sim 0.2$ mag over this redshift range, as found by \citet{Drory+03},
would still be consistent with the data.  For all bins, the data extend to $M_k^\ast$ or
deeper.  To calculate the total luminosity, we will take the integral
of the Schechter function fit; because of the way this is normalized,
this is equivalent to adding up all the observed light brighter than
$M_k^\ast$ (including appropriate statistical weights), and using the
Schechter function to extrapolate to zero 
luminosity.  With this method of normalization the total luminosity is
insensitive to the choice of $M^\ast$, and choosing a weakly evolving
value would have no influence on our results.  However, the integral
does depend on the assumed value of $\alpha$, so we
calculate results for both $\alpha=-1.09$ and $\alpha=-0.8$.  These
results, as well as the total observed luminosity per group (uncorrected for the magnitude
limit), are tabulated in Table~\ref{tab-ltot}. 

The uncertainty in the total luminosity assumes the unweighted number of galaxies
per magnitude bin is described by a Poisson distribution.  This accounts
for the fact that most of the luminosity comes from the few brightest
galaxies, which are generally the least numerous and therefore subject
to the largest statistical fluctuation within a population of similar
groups.  It also
accommodates the possibility that occasionally the brightest group
galaxy may not have a spectroscopic redshift (as long as this galaxy
isn't systematically overlooked in every group, and is not much
brighter than the brightest galaxy for which we do have a redshift).
Furthermore, the uncertainty
scales with the statistical weight on the galaxy, so any incompleteness
at the bright end of the luminosity function will be reflected in
correspondingly larger error bars.  In fact, our spectroscopic completeness is
very high for the brightest galaxies, especially for the R22 sample,
and therefore we are unlikely to have missed the brightest member galaxy in
most groups.  

\begin{figure}
\leavevmode \epsfysize=8cm \epsfbox{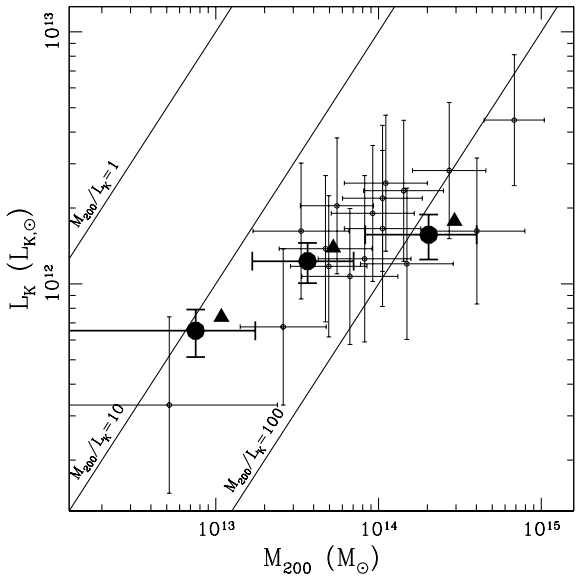} 
\caption{The total near-infrared luminosity of galaxy groups is shown as a function of
  dynamical mass, as estimated from the group velocity dispersion.  The
  small, open symbols are low-redshift data from \citet{RBGMR}.  The
  filled symbols are our groups, in three bins of velocity
  dispersion.  The dynamical mass plotted is the weighted average mass of all
  groups in the bin.  The error bars on this quantity show the minimum
  and maximum mass of the groups in this bin.  The total stellar mass
  includes an extrapolation to zero luminosity, assuming
  $\alpha=-0.8$.  The triangles include a small correction for passive
  evolution, based on a single stellar population model from \citet{BC03}, and for
  hierarchical growth (in both the stellar and dark mass) from the
  Millennium simulation.  The solid lines
  indicate lines of constant \mlk\ ratio.
\label{fig-clusterML}}
\end{figure}
\subsection{Mass-to-light ratios}
We are now in a position to compare the total NIR luminosity
with the dynamical mass of the galaxy groups.  The latter quantity is computed
from the velocity dispersion (Equation~\ref{eqn-vm}) in the same way as \citet{RBGMR}, who performed a
very similar experiment at $z\approx 0$ (see below).  The large, filled circles
in Figure~\ref{fig-clusterML} represent our data, shown in three bins of velocity dispersion, but summed
over all redshifts, as in the last three rows of Table~\ref{tab-ltot}.
The filled triangles demonstrate the effect of applying two, relatively small, evolutionary
corrections to our data.  The first is an estimated correction for passive luminosity
evolution, based on a single stellar population, dust-free \citet{BC03} model, that amounts
to a $\sim30$ per cent reduction in luminosity, at fixed mass.  The
second correction aims to account for the hierarchical mass growth to
$z=0$.  Using the results of the Millennium simulation \citep{Mill_sim}, we find that haloes with
$M>10^{13}M_\odot$ grow by $\sim 43$ per cent
between $z=0.4$ and $z=0$, independent of initial mass.  We assume that this growth
occurs in both the luminosity and mass of the system, so the points
move up and to the right, along lines of constant mass-to-light ratio.
The net result of these two corrections is that the \mlk\ of our
$z=0.4$ groups are best compared to the \mlk\ of $z=0$ systems that are
$\sim$ 40 per cent more massive, as these represent, statistically, the
most likely descendents.

The small, open symbols in Figure~\ref{fig-clusterML} represent the
low-redshift data of \citet{RBGMR}.  This is a sample of groups
selected from a complete spectroscopic survey at $z\approx 0$, 
followed up with subsequent, deeper spectroscopy and cross-correlated
with the 2MASS \citep{2MASS} to obtain
near-infrared magnitudes.  Thus the group selection
is similar to ours, in that it is based on redshift-space clustering
rather than X-ray or radio emission.  There are naturally differences
in our procedure for identifying groups and measuring their velocity
dispersions, in part due to the fact that our galaxies are drawn from an
incomplete redshift survey and, therefore, must be statistically
weighted.  However, due to the small number of members per group, our
velocity dispersions are dominated by statistical uncertainties, rather
than any systematic effects related to exactly which galaxies were
included in the computation.  For example, \citet{CNOC2_groupsI} find
that the deeper, more complete spectroscopic sampling from Magellan
certainly improves the statistical uncertainties of our velocity
dispersions, but does not reveal a systematic bias relative to measurements made
from the original survey.  Since we use the same equations to relate
velocity dispersion to dynamical mass as \citet{RBGMR}, it is, therefore,
possible to compare their data with ours to look for evolutionary trends.

There are two remarkable points about the result in Figure~\ref{fig-clusterML}.  The first is that
our data lie within the scatter of the low-redshift data of
\citet{RBGMR}. There has, therefore, been no strong evolution in the stellar
mass fraction of galaxy groups since $z=0.4$. Indeed, as is evident
from the data in Table~\ref{tab-ltot}, we see no evidence for
significant evolution
within our sample itself, which spans the redshift range $0.1<z<0.6$.  Given the statistical and systematic
uncertainties we can rule out a pure luminosity evolution of a factor $\sim2$ or
greater since $z\sim 0.4$. Also interesting is the suggestion that the \mlk\ ratio
at $z=0.4$ depends on system mass.  The lowest mass groups in our sample
have \mlk$\sim 10$, while the most massive systems are a factor $\sim 10$
larger. Again the large uncertainties preclude strong conclusions, but this
is similar to the trend observed at low-redshift \citep{RBGMR,LMS}.  We will discuss this further
in Section~\ref{sec-discuss}.

\section{Discussion}\label{sec-discuss}
\subsection{Lensing masses}\label{sec-lens}
\begin{table*}
  \centering
    \caption{Similar to Table~\ref{tab-ltot}, but where we have
      included an estimate of the average lensing mass, $\sigma_{\rm lens}$, in
      column (3), for each range of redshift-derived velocity
      dispersions (column 1).  The lensing masses are derived from a fit between the lensing and
      dynamical velocity dispersion estimates in \citet{Parker_thesis},
      and are used to calculate the cluster masses,
      $R_{200}$, and luminosity within $R_{200}$. Their
      uncertainties are the statistical uncertainties given in \citet{Parker_thesis}.
      The other columns
    are as in Table~\ref{tab-ltot}.  The uncertainty in \mlk\ (column
    8) ignores any uncertainty in the mass, which would be correlated
    with the uncertainty in $L_K$.}
    \label{tab-lensmass}
    \begin{tabular}{cccccccc} \hline \hline
(1) & (2) & (3) & (4) & (5) & (6) & (7) & (8)\\ 
  Redshift&$\sigma_{\rm dyn}$    & $\left<\sigma_{\rm lens}\right>$& $L_{K}$, $M_K<M_K^\ast$ &   $L_{K,tot}$  ($\alpha=-0.8$) &  $L_{K,tot}$ ($\alpha=-1.09$) &   $M_{200}$ & $M_{200}/L_{K,tot}$\\
   & (\kms)      &  (estimated)&($10^{11} L_{K,\odot}$)           &             ($10^{11} L_{K,\odot}$)  &  ($10^{11} L_{K,\odot}$)   & ($10^{13} M_\odot)$& ($M_{\odot}/L_{K,\odot})$\\
\hline
0.1--0.6 & 0--250   & $193\pm  38$  & $ 2.81\pm0.6$ & $6.19\pm1.3$ & $8.53\pm1.8$ &   0.88 & $14.15\pm3$\\
         & 250--425 & $270\pm  39$  & $ 5.57\pm1$ & $12.3\pm2.2$ & $16.95\pm3.1$ &   2.08 & $16.93\pm3.1$\\
         & 425--800 & $425\pm 182$  & $ 7.02\pm1.4$ & $15.5\pm3.1$ & $21.35\pm4.3$ &   6.53 & $42.1\pm8.5$\\
\hline
    \end{tabular}
\end{table*}
Dynamical mass estimates based on velocity dispersion can be
problematic for groups, both because of the small number of galaxies
involved, and because of the difficulty of testing the necessary
assumptions required to link velocity dispersion to mass.  It is
unclear, for example, how close these groups are to dynamical equilibrium.  Weak lensing
provides a promising alternative, as this method measures the mass
independently of the dynamical state of the group.  Of course, this method
has its own drawbacks, which are mainly the weakness of the signal, the
sensitivity to choice of group centre, and the
sensitivity to all mass projected along the line of sight between the
observer and the lensed galaxy.  Nonetheless, as the sources of systematic
uncertainty are quite different from those associated with the
velocity dispersions, they provide a useful check on our results.

Weak lensing masses have been measured for the CNOC2 group sample by \citet{Hoek01}
and \citet{PHCH}.  However, due to the weakness of the lensing signal,
this can only be done for ensembles, and  \citet{Parker_thesis}
provide masses for three classes of groups, split by the velocity
dispersion \citep[as measured by ][]{CNOC_groups} at $\sigma=190$ \kms\ and 
$\sigma=500$ \kms.  We make a linear fit to relate the lensing
dispersions $\sigma_{\rm lens}$ (the velocity dispersion associated
with an isothermal sphere of the measured mass) to the average
dynamical values $\sigma_{\rm dyn}$, and find\footnote{To be fully consistent, we use the original measurements of
$\sigma_{\rm dyn}$ from \citet{CNOC_groups} to bin the groups in the
same way as \citet{PHCH}, but then use our updated values to compute
the average $\sigma_{\rm dyn}$ within each bin.} $\sigma_{\rm lens}=113\mbox{km~s}^{-1}+0.49\sigma_{\rm dyn}$.
Using  this fit as an estimate of the lensing mass for each individual group, we recompute $R_{200}$ and
$M_{\rm 200}$ and repeat the entire analysis to measure the
corresponding total NIR luminoisty.  This includes
recomputing the NIR completeness of each group, since this depends on
$R_{200}$, as well as the total luminosity within $R_{200}$.  The
results are shown in Table~\ref{tab-lensmass}. 

The main difference from the results in \S~\ref{sec-lgroup} is that the
masses of the groups with the highest velocity dispersion are
lower than estimated using Equation~\ref{eqn-vm}.  This leads to a
lower value of \mlk\ (despite the corresponding reduction in
$R_{200}$) and reduces the increase of \mlk\ with system
mass to a factor $\sim 3$ between the lowest- and highest-mass groups in our
sample.  

\subsection{Stellar mass-to-light ratios}\label{sec-stellarml}
While the near-infrared luminosities are closely related to stellar
mass, there is still substantial variation in the stellar mass-to-light
ratio (\mls)
with galaxy colour.  We use the \citet{BC03} models to compute
\mls\ as a function of galaxy colour, for a range of
parameters.  Assuming a \citet{Chab} initial mass function, we model
blue galaxies ($B-V<0.4$) as young galaxies with constant star formation
and dust extinction of $\tau_v=1$ mag; this gives \mls$=0.2$.
For red galaxies ($B-V>1$), a 11.7 Gyr old, dust-free single stellar
population model gives \mls$=0.7$.  For intermediate colours, we
simply adopt a linear interpolation between these values.
\begin{figure}
\leavevmode \epsfysize=8cm \epsfbox{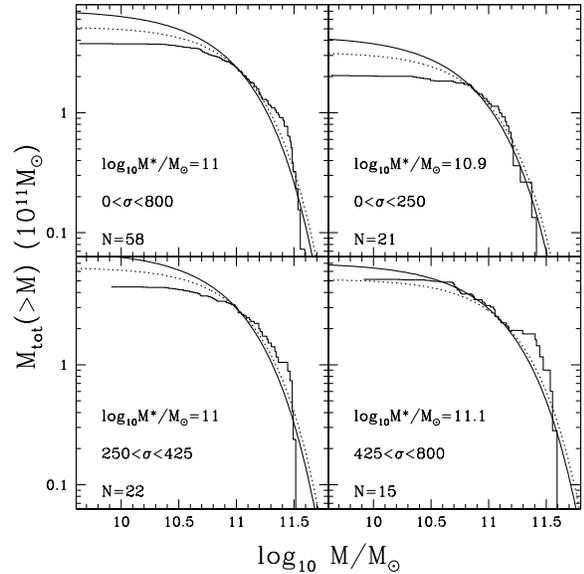}
\caption{The cumulative stellar mass function for the whole group sample (top left
  panel) and for samples subdivided into bins of velocity dispersion (remaining
  panels, as labeled).  The stellar masses are computed from the
  $K_{rest}$ values, assuming the simple, colour-dependent \mls\
  model described in the text.
  Each panel shows the number of groups contributing to each bin, and
  the value of $M^\ast$ assumed for the Schechter functions, plotted as
  curved lines.  The solid line assumes a faint end slope of
  $\alpha=-1.09$, while the dotted line assumes $\alpha=-0.8$.  The
  Schechter functions are normalized to match the observed total stellar mass in
  galaxies with $M>M^\ast$.
\label{fig-mfunc}}
\end{figure}
\begin{figure}
\leavevmode \epsfysize=8cm \epsfbox{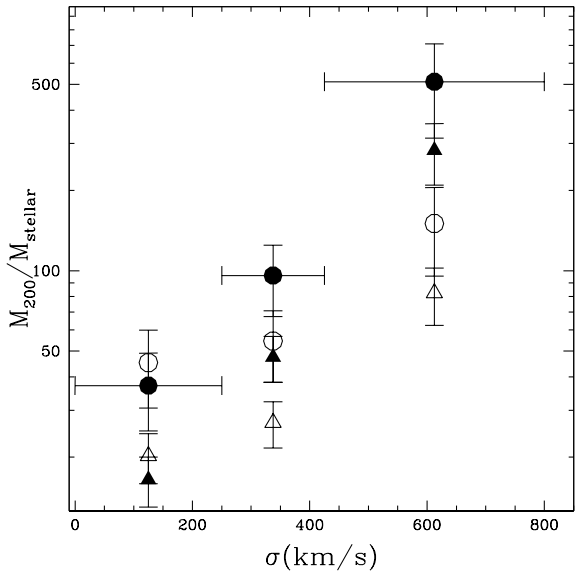}
\caption{The dynamical-to-stellar mass ratio as a function of velocity
  dispersion.  The solid points are computed assuming dynamical masses and
  $R_{200}$ derived from the velocity distribution, while the open
  symbols use the weak-lensing derived masses.  The circles are based
  on our simple colour-dependent 
  \mls\ model, while the triangles assume \mls$=0.7$ for all galaxies.
  The three velocity dispersion bins are the same as in
  Figure~\ref{fig-mfunc}, and the points are shown in the middle of
  each bin with the range indicated by the error bars on one set of
  points only, for clarity.  
\label{fig-mdms}}
\end{figure}

The cumulative stellar mass functions are shown in Figure~\ref{fig-mfunc}, for the
whole group sample and for subsamples divided by velocity dispersion.
The mass functions can be reasonably approximated by the Schechter
functions shown, which have characteristic mass scales ranging from
$10^{10.9}M_\odot$ for the lowest-mass groups, and $10^{11.1}M_\odot$
for the highest-mass groups.  
This trend for $M^\ast$ to increase with
density is well-known \citep[e.g.][]{BaldryV}.  
For the
entire group sample, the ratio of dynamical mass to total stellar mass
is $168\pm25$.
In Figure~\ref{fig-mdms}
we show the dynamical-to-stellar mass ratio as a 
function of velocity dispersion, for
both the $\sigma-$based and weak lensing-based mass (and
$R_{200}$) estimates.  We also show the effect\footnote{The difference
between these two measurements is
not simply related to the difference in average \mls, because
the fixed value of $M^\ast$ means the normalization of the Schechter
function, and hence the extrapolation to $L_k=0$, also differs in the
two cases.}
of using a constant \mls$=0.7$, as expected for early type galaxies, and
typical of the Universal average at low redshift
\citep[e.g.][]{Cole-2mass_short}.  The spread in the points at a given
velocity dispersion therefore gives
some indication of the systematic uncertainty on the measured $M_{\rm 200}/M_{\rm
  stellar}$.
As already noted, there is some discrepancy between the
lensing-based and redshift-based results for the systems with the
highest velocity dispersions, but overall the data show that $M_{\rm 200}/M_{\rm
  stellar}$ increases from $\lesssim 50$ in the poorest systems to
$\gtrsim100$ in the massive groups.  This could be interpreted as a factor $\sim 2$
difference in star formation efficiency between these haloes.

\subsection{Colours and emission line strengths}
In \citet{CNOC2_groupsI}, we found that the fraction of galaxies with
\oii\ emission increased with decreasing, rest-frame $B_J$-band luminosity,
for both the group and field populations in the CNOC2 (R22) galaxy sample.    We also found that, at
fixed luminosity, emission line galaxies were more common in the field
population.  Using toy models \citep{CNOC2_groupsII} we claimed that this difference was due
to a recent truncation of star formation in group galaxies,
qualitatively similar to what is observed in more massive clusters at
this redshift \citep[e.g.][]{B+97,lowlx-spectra_short,Nakata,Pogg05}. However,
that analysis was complicated by the fact that the rest-frame $B_J$-band
luminosity is sensitive to recent star formation.  Here we can
re-examine the correlation as a function of stellar mass.

\begin{figure}
\leavevmode \epsfysize=8cm \epsfbox{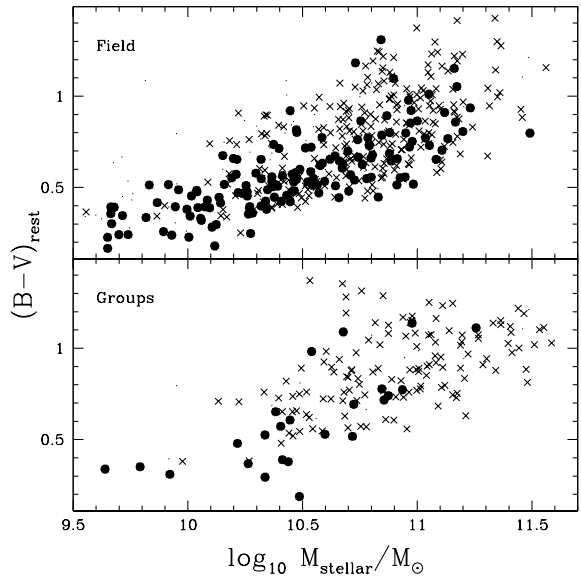} 
\caption{The correlation between (B-V) rest-frame colour and stellar mass, for the group (bottom panel)
  and field (top panel) samples.  Both samples are restricted to the
  redshift range $0.12<z<0.55$, where there is no colour incompleteness.  Only galaxies brighter than $K=17.2$
  (corresponding to $R=21.5$)  are
  shown (crosses), and group galaxies are restricted to those within $R_{200}$.  Filled circles represent galaxies
  with strong emission lines, \ewoii$>10$\AA.  The few small points
  represent galaxies for which no \ewoii\ measurement is available.
\label{fig-cmass}}
\end{figure}

Figure~\ref{fig-cmass} shows the rest-frame (B-V) colour as a function
of stellar mass. We restrict the redshift range to 
$0.12<z<0.55$; outside this range, colour-dependent incompleteness
becomes important because few identifiable absorption features of red
galaxies lie within the band-limiting
filter used for the original CNOC2 spectroscopy \citep{CNOC2-I}.  Within this
range, all galaxies
with redshifts are significantly detected in all filters, so there is
no colour bias.  Galaxies with strong
emission lines (\ewoii$>10$\AA) are shown as filled symbols. 

\begin{figure}
\leavevmode \epsfysize=8cm \epsfbox{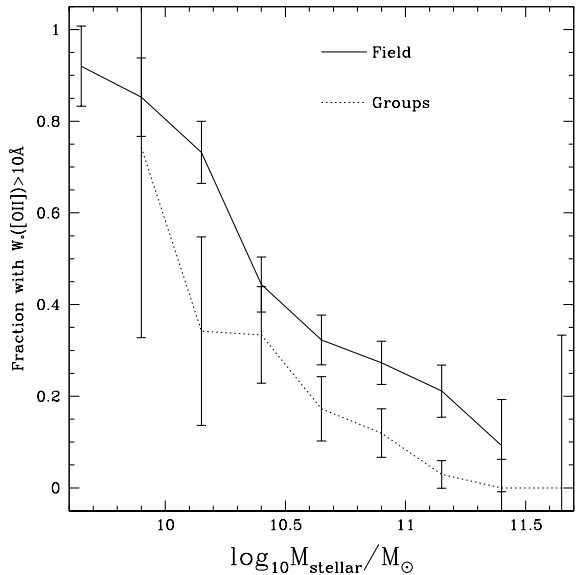} 
\caption{The fraction of galaxies with strong \oii\ emission lines is
  shown as a function of stellar mass, for the field and group sample
  at $0.12<z<0.55$.  Error bars are 1-$\sigma$ jackknife resampling estimates.
\label{fig-cmass3}}
\end{figure}

Both the group and field populations exhibit a decrease in the fraction of blue,
emission line galaxies as stellar mass increases.  However, at fixed
stellar mass, such late-type galaxies are less common in the group
environment, compared with the field.  This is illustrated more clearly
in Figure~\ref{fig-cmass3}, where 
we show the fraction of galaxies with strong
emission lines (\ewoii$>10$\AA) as a function of stellar mass, for the
field and group samples.  Thus we confirm our conclusions in
\citet{CNOC2_groupsII}, that star formation in group environments was
terminated earlier than in the lower-density field at $z\sim0.4$.

\subsection{Interpretation}
The dominance of old stellar populations, together with the
weakly-evolving mass-to-light ratio, implies that most of the stellar
mass in groups must have already been in place by $z\sim 0.4$. 
Thus, we see no evidence of accelerated star formation in groups,
despite the fact that the low velocity dispersions are expected to lead to enhanced
dynamical interactions between group members.  This is especially
surprising since our sample is at an epoch when
the global star formation rate was several times higher than it is at
present \citep[e.g.][]{Hopkins04}.  Unlike with group samples selected from their
X-ray emission \citep[e.g.][]{MLFRJ}, we have little reason to
expect our redshift-selected sample to be biased toward especially relaxed or evolved
groups, and therefore conclude that older stellar populations are
characteristic of groups at $z\sim 0.4$ in general.   

Recent numerical simulations suggest that in overdense environments,
dark matter haloes assembled their mass more rapidly,
and at higher redshift, than haloes of the same mass in low-density
environments \citep{GSW}, and this might be sufficient to explain the more evolved
populations of galaxy groups and clusters without appealing to local
interactions between galaxies and their environment \citep{M+06}.
Determining whether or not this is a viable explanation for the observations
presented here awaits a quantitative comparison with theoretical
models.

\section{Conclusions}\label{sec-conc}
We have presented near-infrared observations of 58 redshift-selected
galaxy groups at $0.1<z<0.6$, obtained from the WHT and {\it Spitzer}
space telescope. This affords us the first opportunity to obtain reliable
stellar masses for galaxies in groups at this redshift.  From these
data we draw the following conclusions:

\begin{itemize}
\item There is evidence that the highest-mass groups in our sample
  ($\sigma>500$~\kms) have
  \mlk\ that are a factor $\gtrsim 3$ larger than for the 
  lowest-mass groups ($\sigma<200$~\kms).  This
  trend is present whether we use dynamically-estimated masses (from
  the velocity dispersions) or masses measured from a weak lensing analysis.
\item Our best estimate of the group stellar mass fraction
  decreases from $\sim 2$ per cent at $\sigma<200$ \kms, to $\lesssim
  1$ per cent in the most massive groups with $\sigma>500$ \kms.
\item When comparing groups at $z\sim0.4$ to their statistically most-likely descendants
  at $z=0$, we find no evidence for strong evolution in \mlk\ beyond
  that expected for a passively evolving population.  
\item Group galaxies have older stellar populations, as measured by
  their \ewoii\ emission or rest-frame (B-V) colour, than field galaxies of the same
  stellar mass.   
\end{itemize}

We have demonstrated that galaxies in groups are
predominantly old, passively evolving systems, even at $z\sim 0.4$.
This is something of a surprise, given that
effects such as tidal interactions and mergers (which are expected to
induce star formation) should be more common, while more dramatic
effects like ram-pressure stripping, often thought to play a role in
terminating star formation within more massive clusters, should be
negligible.  Our results imply that these effects have not had a large
influence on the properties of galaxies in groups since
$z\sim 0.4$.

\section{Acknowledgments}\label{sec-akn}
We would like to thank the CNOC2 team for allowing us access to their
unpublished data.  We also gratefully acknowledge the efforts of the
Virgo consortium, who have very usefully made the Millennium simulation
results easily accessible, and we thank them specifically for allowing
us access to the full simulation results prior to public release.  M. Balogh would like to extend an especially warm
thanks to Robert Greimel who completed the INGRID observations
following a telescope scheduling change.
This research was supported by the Natural Sciences and Engineering
Research Council of Canada, through a Discovery Grant to M. Balogh, and
an Undergraduate Research award to R. Henderson.  D. Wilman is supported by a Max
Planck Society Postdoctoral Fellowship. R. Henderson would
like to thank the astronomy group at Waterloo, especially Gretchen
Harris, for many helpful discussions.  
\bibliography{ms}
\end{document}